\documentclass[twocolumn,showpacs,pra,aps,superscriptaddress]{revtex4-1}
\usepackage{amsmath}
\usepackage{amsfonts}
\usepackage{amssymb}
\usepackage{graphicx}
\usepackage{hyperref}
\usepackage{bm}
\usepackage{color}
\usepackage{siunitx}
\hypersetup{backref,
colorlinks=true,
linkcolor=cyan,
linktoc=page,
citecolor=cyan}

\usepackage{tikz}
\usetikzlibrary{arrows}
\usepackage{pgfplots}
\usepackage{graphicx}

\hypersetup{backref,
colorlinks=true,
linkcolor=blue,
linktoc=black,
citecolor=cyan,
urlcolor=denim}

\newcommand{\ketp}[1]{$\left|#1\right\rangle$}
\definecolor{Green}{cmyk}{1,0.,1,0}
\newcommand{\pr}[1]{\ensuremath{\left[#1\right]}} 
\newcommand{\pc}[1]{\ensuremath{\left(#1\right)}}

\newcommand{\ket}[1]{\ensuremath{\left\vert#1\right\rangle}}

\newcommand{\rec}{\ensuremath{_\text{rec}}}

\def\XXint#1#2#3{{\setbox0=\hbox{$#1{#2#3}{\int}$}
     \vcenter{\hbox{$#2#3$}}\kern-.5\wd0}}

\renewcommand{\Re}{\operatorname{Re}}
\renewcommand{\Im}{\operatorname{Im}}

\definecolor{burgundy}{rgb}{0.5, 0.0, 0.13}
\definecolor{denim}{rgb}{0.08, 0.38, 0.74}
\definecolor{midnightgreen}{rgb}{0.0, 0.29, 0.33}
\definecolor{sienna}{rgb}{0.53, 0.18, 0.09}
\definecolor{sacramentostategreen}{rgb}{0.0, 0.34, 0.25}

\newcommand{\ev}{\ensuremath{\text{ev.}}}
\newcommand{\kb}{\ensuremath{k_\text{B}}}
\newcommand{\ct}{\ensuremath{\tilde{c}}}
\newcommand{\xit}{\ensuremath{\tilde{\xi}}}
\newcommand{\kt}{\ensuremath{\tilde{\kappa}}}

\begin{document}
\title{
Radio-frequency evaporation in an optical dipole trap
}

\author{Raphael Lopes}
\address{Laboratoire Kastler Brossel, Coll\`ege de France, CNRS, ENS-PSL Research University, Sorbonne Universit\'e, 11 place Marcelin Berthelot, 75005 Paris, France}

\begin{abstract}
We present an evaporative cooling technique for atoms trapped in an optical dipole trap that benefits from narrow optical transitions. For an appropriate choice of wavelength and polarization, a single laser beam leads to opposite light-shifts in two internal states of the lowest energy manifold. Radio-frequency coupling between these two states results in evaporative cooling at a constant trap stiffness. The evaporation protocol is well adapted to several atomic species, in particular to the case of Lanthanides such as Er, Dy, and fermionic Yb, but also to alkali-earth metals such as fermionic Sr. We derive the dimensionless expressions that allow us to estimate the evaporation efficiency. As a concrete example, we consider the case of $^{162}$Dy and present a numerical analysis of the evaporation in a dipole trap near the $J'=J$ optical transition at \SI{832}{\nano\meter}. We show that this technique can lead to runaway evaporation in a minimalist experimental setup. 
\end{abstract}
\maketitle

A key step to achieve Bose-Einstein condensation \cite{Anderson1995} is the evaporative cooling technique introduced in the 90s \cite{Hess1986, Luiten1996a, Ketterle1996, Guery-Odelin1998}. Such a mechanism is qualitatively simple to understand; particles with high energy, above a given cut-off energy, $\epsilon_\text{c}$, are lost from the system followed by subsequent thermalization. The truncated Boltzmann distribution readjusts, and the temperature falls at the cost of particle loss. 

This process has proved successful in cold atom experiments, in both magnetic and optical traps \cite{Adams1997, Inguscio1999, Ketterle2002, Cornell2002, Colombe2004}. In magnetic traps, a magnetic field gradient ensures that atoms prepared in a low-field seeking internal state are trapped while atoms in a high-field seeking state are expelled. The coupling between the two states is ensured by a radio-frequency (RF) photon which frequency is progressively swept to reduce the cut-off energy. For efficient evaporative cooling to occur the elastic collision rate, $\Gamma_\text{el.}$, needs to dominate over the loss rate, $\Gamma_\text{loss}$. In that case, evaporative cooling accelerates over time, reaching degeneracy at the cost of minimal particle loss.

Nowadays, most cold atom experiments use evaporative cooling in optical dipole traps as it allows the cooling of different internal states and species with zero magnetic moment \cite{Grimm2000, Barrett2001b, Granade2002, Takasu2003, Weber2003, Fukuhara2007a}. However, it comes at the cost of reduced evaporation efficiency, as the evaporation is performed by continuously decreasing the trap depth, which softens the potential and reduces the elastic collision rate \cite{OHara2001}. More elaborated strategies, such as addressing resonant optical transitions \cite{Wilkowski2010}, combining dipole traps with very different volumes \cite{Kinoshita2005, Clement2009},  or changing the s-wave scattering length, $a$, \cite{Weber2003} during the evaporation, allow to mitigate this issue but lead to an enhanced experimental complexity and secondary inelastic processes~\cite{Chin2010a}.

\begin{figure}
\centerline{
  \resizebox{9cm}{!}{
    \begin{tikzpicture}[
      scale=0.5
    ]
    \draw[line width=0.25mm, black ] (0,-1) -- (2,-1) node [right] {\ketp{d}};
    \draw[line width=0.4mm, gray ] (4,-0.5) -- (6,-0.5) node [right] {\ketp{b}};
    \draw[line width=0.8mm, burgundy ] (4,4) -- (6,4) node [right, xshift=0.1cm] {\ketp{e_1}};
    \draw[line width=1.4mm, denim ] (0,7) -- (2,7) ;
    \draw[line width=1.4mm, denim ] (4,7.5) -- (6,7.5) node [right, yshift=-0.2cm, xshift=0.1cm] {\ketp{e_2}};
    \draw[line width=0.3mm, densely dashed, burgundy] (4,4.7) -- (6,4.7) ;
    \draw[line width=0.5mm, densely dashed,denim] (0,4.2) -- (2,4.2) ;
    \draw[line width=0.3mm,<->,shorten >=2pt,shorten <=2pt,>=stealth] (0.5,4.2) -- (00.5,7) node[midway,left] {$\Delta$};
    \draw[line width=0.3mm,>-<,shorten >=-3pt,shorten <=-3pt,>=stealth] (4.2,4.) -- (4.2,4.7) node[midway,left] {$\delta$};
    \draw[line width=0.5mm,->,shorten >=2pt,shorten <=0.5pt,>=stealth,burgundy] (1,-1) -- (1,4.2);
    \draw[line width=0.5mm,->,shorten >=2pt,shorten <=0.5pt,>=stealth,burgundy] (5.,-0.5) -- (5.,4.8) node[midway, left]{$\lambda_1$};
   \draw[line width = 0.2 mm,double,<->,shorten >=4pt,shorten <=4pt,>=stealth, sacramentostategreen!80] (11.5,0.8) -- (11.5,3.2);
    \begin{axis}[
	 name=one,
	axis y line =center,
	axis x line =bottom,
	axis line style ={->, line width = 0.5mm},
	width=7cm,
	height=7cm,
	scale only axis,
	xmin=-1,xmax=1,
	ymin=-0.5,ymax=1.4,
	ytick={-5},
	xtick={-5},
	xlabel={},
	xlabel style={font=\huge, yshift=-1mm},
	yticklabel style={xshift=0.5mm},
	tick label style = {font=\huge},
	tick label style = {font=\huge},
	at={(9cm,-0.5cm)}, anchor=south west
        ] 
    \addplot [domain=-1:1, samples=100, ultra thick, gray] {0.2+exp(-(x^2)/0.06)};
    \addplot [domain=-1:1, samples=100, ultra thick, black] {0.0-0.25*exp(-(x^2)/0.06)};
    \end{axis}
    \node[at=(one.left of north west), anchor= west,yshift=-2.9cm,xshift=2.7cm,black] {\ketp{d}};
    \node[at=(one.left of north west), anchor= west,yshift=-1.9cm,xshift=2.7cm,gray] {\ketp{b}};
    \node[at=(one.left of north west), anchor= west,yshift=0.3cm,xshift=1.5cm] {$V$ (a.u.)};
    \node[at=(one.left of north west), anchor= west,yshift=-3.8cm,xshift=2.45cm] {$r$ (a.u.)};
    \node[at=(one.left of north west), anchor= west,yshift=-2.38cm,xshift=00.3cm,sacramentostategreen!80] {RF};
    \node[at=(one.left of north west), anchor= west,xshift=-4.70cm,yshift=0.8cm] {(a)};
    \node[at=(one.left of north west), anchor= west,xshift=-0.00cm,yshift=0.8cm] {(b)};
     \end{tikzpicture}
  }
}
\caption{Schematic representation of RF-evaporation in an optical dipole trap. (a) Schematic energy representation of the two lowest energy states $\ket{d}$ and $\ket{b}$, and excited states $\ket{e_1}$ and $\ket{e_2}$. A laser beam with wavelength $\lambda_1$ induces opposite light-shifts for the two states. (b) Spatial dependence of the optical potential for the dark-state $\ket{d}$ and bright-state $\ket{b}$ for $\Delta/\delta<0$. Radio-frequency couples the two states and defines the cut-off energy for atoms in $\ket{d}$.}
\label{figRFevap}
\end{figure}
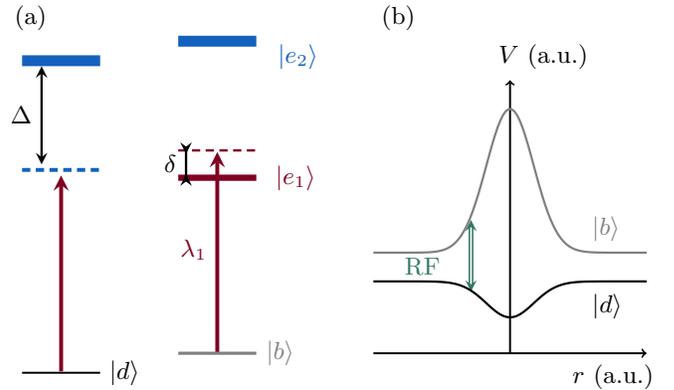

In here, we propose to cool a sample trapped in a single gaussian laser beam, with wavelength $\lambda_1$, close to a narrow optical transition (see Fig.~\ref{figRFevap}). The two lowest energy states, $\ket{d}$ and $\ket{b}$, can, for a given polarization, feel opposite light-shifts. For instance, if $\ket{d}$ is not coupled to $\ket{e_1}$ (see Fig.~\ref{figRFevap}), its polarizability is set by other far-detuned excited states, that we regroup under $\ket{e_2}$, with detuning $\Delta$. On the other hand, the polarizability of the bright-state $\ket{b}$ is, to a good approximation, simply defined by the detuning from the narrow optical transition, $\delta$, and the transition linewidth $\Gamma$. For $\delta>0$, $\ket{b}$ has a negative polarizability and therefore the atoms feel a repulsive potential at the intensity maximum, while atoms in $\ket{d}$ are trapped for $\Delta<0$. Coupling the two internal states, for instant through radio-frequency, results in evaporative cooling in an optical trap with a fixed trap stiffness, where $\epsilon_\text{c}$ is set by the RF-frequency. Importantly, the narrowness of the optical transition ensures that atoms in $\ket{b}$ are not repumped into $\ket{d}$ before escaping the laser beam spatial profile. 

This technique is experimentally simple to implement as it only requires a single, tighly-focused, laser beam overlapping with the atomic cloud.  
Several advantages of optical dipole traps are still applicable, such as the possibility to cool two distinct internal states, in the case of two dark-states, sympathetic cooling of atomic mixtures, and evaporative cooling of states with zero magnetic moment. Moreover, the same protocol can be extended to box-like potentials \cite{Gaunt2013}. Compared to the cooling technique reported in Ref.~\cite{Wilkowski2010}, our method is experimentally less demanding as we do not require a long-lived metastable state nor anti-magic wavelengths.

\section{Applicability to different atomic species}

In summary, two main ingredients are required to perform the discussed radio-frequency evaporation in an optical dipole trap. A ground level with at least two internal states and a narrow optical transition with respect to which only one of the two aforementioned states can be coupled to.

In that sense, Erbium and Dysprosium atomic species are excellent candidates, since several narrow optical transitions exist and the background polarizability is positive for $\lambda \gtrsim \SI{400}{\nano\meter}$ \cite{Li2017, Becher2018}.  For Dysprosium, an interesting choice is the $J'=J$ transition with wavelength $\lambda \approx \SI{832}{\nano\meter}$, and linewidth $\Gamma \approx 2\pi \times \SI{15}{\kilo\hertz} $. At this wavelength, the imaginary part of the background polarizability is small, and heating is negligible. A similar transition exists for Erbium around \SI{847.5}{\nano\meter} \cite{Becher2018}.
Other species can also benefit from this evaporation protocol, such as fermionic Strontium and Ytterbium near the narrow optical transitions $F\rightarrow F'=F$ at \SI{689}{\nano\meter} and \SI{556}{\nano\meter}, respectively \cite{Stellmer2013, Fukuhara2007a}, or in the case of Titanium, using the $J'=J=4$ transition at \SI{546}{\nano\meter} with linewidth $\Gamma \approx 2\pi \times \SI{300}{\kilo\hertz}$ \cite{Eustice2020}.

\section{Rate equations}

In order to investigate the evaporation efficiency we consider the truncated Boltzmann distribution approximation, and write the instantaneous time variation of the atom number, $N$, total energy, $E$, and temperature, $T$ \cite{Luiten1996a,Guery-Odelin1998},
\begin{align}
\label{eq:ratechange1}
\dot{N} &= \dot{N}_\ev + \dot{N}_\text{spl.} + \dot{N}_\text{loss.}  \\
\dot{E} &= \dot{E}_\ev + \dot{E}_\text{spl.} + \dot{E}_\text{loss.}  +\dot{E}_\text{heat.}  \\
\dot{T}/T & = \dot{E}/E - \dot{N}/N   \, ,
\label{eq:ratechange3}
\end{align}
where $N_\ev$ and $E_\ev$ are associated to atom loss and energy reduction through evaporation, $N_\text{spl.}$ and $E_\text{spl.}$, to particle spilling from the trap, $N_\text{loss}$ and $E_\text{loss}$ one-body loss through collisions with the residual background gas, and $E_\text{heat.}$ heating induced by incoherent photon scattering processes (see below).
The internal energy is given by $E = N \kb T \tilde{c}$, with $\ct= \text{dLog}(\xi) / \text{dLog}T$, and
$\xi =\pc{1/n \lambda^3 } \int_0^\infty  \text{d} \epsilon \, \rho(\epsilon) f(\epsilon) $,  where
\begin{align}
\rho (\epsilon) = \frac{2\pi (2m)^{3/2}}{(2\pi \hbar)^3} \int_{U(r)\leq \epsilon} \text{d}^3\textbf{r}\, \sqrt{\epsilon - U(\mathbf{r})}\, ,
\end{align}
is the energy density of states, $f(\epsilon ) = n \lambda^3 e^{-\beta \epsilon} \Theta(\epsilon_\text{c}-\epsilon) $ the truncated Boltzmann distribution, $U(\mathbf{r})$ the conservative potential, $n$ the density, $m$ the atomic mass, $\lambda$ the deBroglie wavelength, $\beta = 1/\kb T$, $\kb$ the Boltzmann constant, $\hbar$ the reduced Planck constant, and $\Theta(\epsilon)$ the Heaviside function. 

Four processes define the evolution of temperature and atom number \cite{Luiten1996a, Ketterle1996, Guery-Odelin1998, Yan2011}. The loss of particles through evaporation leads to changes of atom number and total energy,
\begin{align}
\label{eqNev}
\dot{N}_\ev/N &= - \Gamma_\ev \, n \sigma v \\
\dot{E}_\ev/E &= - \frac{\Gamma_\ev}{\ct} \kt  \, n \sigma v\,,
\label{eqEev}
\end{align}
with evaporation rate $\Gamma_\ev = e^{-\eta} \, V_\ev/V_\text{e}$, and volumes
\begin{align*}
V_\text{e} & = N/n =  \lambda^3 \int_0^{\epsilon_c}  \text{d}\epsilon\,\rho(\epsilon) e^{-\beta\epsilon} \\
V_\ev &= \lambda^3 \beta \int_0^{\epsilon_\text{c}} \text{d}\epsilon\, \rho (\epsilon) \pr{\pc{\epsilon_\text{c}- \epsilon - \kb T}e^{-\epsilon \beta} + \kb T e^{-\eta}} \\
X_\ev &= \lambda^3 \beta \int_0^{\epsilon_\text{c}} \text{d}\epsilon\, \rho (\epsilon) \pr{\kb T e^{-\epsilon \beta} - \pc{\epsilon_\text{c} + \kb T -\epsilon}e^{-\eta}} \, ,
\end{align*}
where $\kt = \pc{\eta + 1 - X_\ev / V_\ev}$ and $\eta = \epsilon_\text{c}/ \kb T$.
The elastic cross section is given by $\sigma = 8\pi a^2$, and the averaged speed by $v = \sqrt{{8 \kb T}/{\pi m}}$.

Spilling also occurs as the cut-off energy is progressively changed. This mechanism induces losses
\begin{align}
\dot{N}_\text{spl.}/N  &=  \xit \, \frac{\dot{T}}{T} \, ,
\label{eqNspl}
\end{align}
where $\xit = e^{-\eta} \rho(\epsilon_\text{c}) \epsilon_\text{c} / \xi$, and an energy change
\begin{align}
\dot{E}_\text{spl.}/E &=   \frac{\eta \xit}{\ct} \, \frac{\dot{T}}{T}\, .
\label{eqEspl}
\end{align}
High-energy collisions with the residual background gas lead to the atom number reduction 
\begin{align}
\dot{N}_\text{loss}/N = -\Gamma_\text{loss}\, ,
\label{eqNloss}
\end{align} with one-body loss rate, $\Gamma_\text{loss}$, and to an energy change 
\begin{align}
\dot{E}_\text{loss}/E = -\Gamma_\text{loss}\, ,
\label{eqEloss}
\end{align}
without heating.

Finally, we also take into account the residual heating associated with incoherent light scattering processes induced by the laser beam, which increases the total energy 
\begin{align}
\dot{E}_\text{heat.}/E =   \pc{Q/\ct \, \kb T} \, ,
\label{eqEheat}
\end{align}
where $Q = 2 U_0 \,\omega\rec/\mathfrak{m}$, $U_0$ the potential depth, $\alpha$ the polarizability, $\mathfrak{m} =\Re\pr{\alpha}/\Im\pr{\alpha}$, and $\hbar \omega\rec$ the recoil energy. We assume a constant value of $Q$ and neglect particle losses resulting from this process.

\begin{figure}[t!]
\begin{center}
\includegraphics[width=0.8\columnwidth]{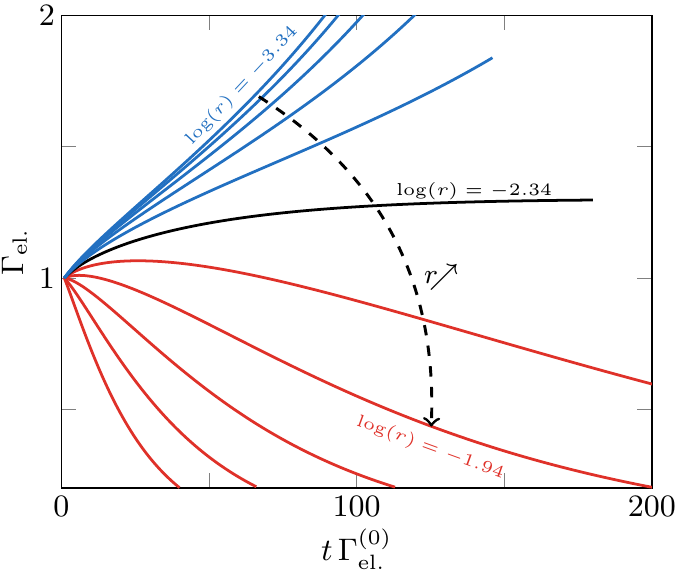}
\caption{{Evolution of the normalized elastic collision rate for $\tilde{U}_0 = 1.25$, $\eta=5.8$ and $s=10^{-7}$. The blue curves correspond to values of $r$ for which the elastic collision rate increases monotonically with time until a gain of three decades is reached in phase-space density. For values of $r>10^{-2.34}$ this is no longer the case (red). The black line represents the elastic collision rate evolution for the limiting case.}\label{fig2a} }
\end{center}
\end{figure} 

Efficient evaporation is characterized by an increasing elastic collision rate, $\Gamma_\text{el.}= n \sigma v$, which accelerates the sample's thermalization, leading to runaway evaporation. Combining Eqs.~\eqref{eq:ratechange1}-\eqref{eq:ratechange3} with Eqs.~\eqref{eqNev}-\eqref{eqEheat}, and assuming a constant value of $\eta$, we express the density and velocity time evolution,
\begin{align*}
\frac{\dot{n}}{n}  &= \frac{\dot{N}}{N} - \ct_2 \frac{\dot{T}}{T} =  -\Gamma_\text{loss} + \Gamma_\ev \, n v \sigma\, \pr{\frac{(\ct_2 - \xit)(\kt - \ct)}{\ct - \xit(\eta - \ct)} -1}  \nonumber \\ &+  \frac{Q}{\kb T} \frac{\xit - \ct_2}{\ct - \xit(\eta - \ct)} \\
\frac{\dot{v}}{v} & = \frac{1}{2}\frac{\dot{T}}{T} =  -\frac{\Gamma_\ev}{2} \, n v \sigma \, \frac{\kt -\ct }{\ct - \xit (\eta - \ct)} + \frac{Q}{\kb T} \frac{1}{\ct - \xit (\eta -\ct)}\, ,
\end{align*}
with $\ct_2 = \ct + \xit -3/2  $. These expressions can be simplified by normalizing the density and averaged speed by their respective values at t=0, $n\rightarrow n / n_0$, and $v\rightarrow v/v_0$. The time is normalized by the initial collision timescale, $t\rightarrow t\, \Gamma_\text{el.}^{(0)}$ where $ \Gamma_\text{el.}^{(0)} = n_0  \sigma v_0$. The coupled equations are then given by 
\begin{align}
\label{eqn}
\dot{n}/n &= -r +A (\eta,\, v)\, n v - \tilde{U}_0B(\eta,\, v)\,  \frac{s}{v^2}\\
\dot{v}/v &= -C(\eta,\, v)\, n v+ \tilde{U}_0 D(\eta,\, v)\, \frac{s}{v^2} \, ,
\label{eqv}
\end{align}
with $\tilde{U}_0~=~U_0 / \epsilon_\text{c}~(t=0)$, and $A(\eta,v),\, B(\eta,v),\, C(\eta,v)$, $D(\eta,v)$ dimensionless, positive, functions which tend to a constant value when $v\rightarrow 0$ (see Appendix). The dimensionless quantities $r = \Gamma_\text{loss}/\Gamma_\text{el.}^{(0)}$ and $s =\frac{1}{\mathfrak{m}}\pc{\omega\rec/\Gamma_\text{el.}^{(0)}}$ relate to experimental limitations. In detail, $r$ characterizes one-body losses, and $s$ heating due to spontaneous emission. A larger $s$ means a larger heating and a larger $r$ more losses.

\begin{figure}[t!]
\begin{center}
\includegraphics[width=\columnwidth]{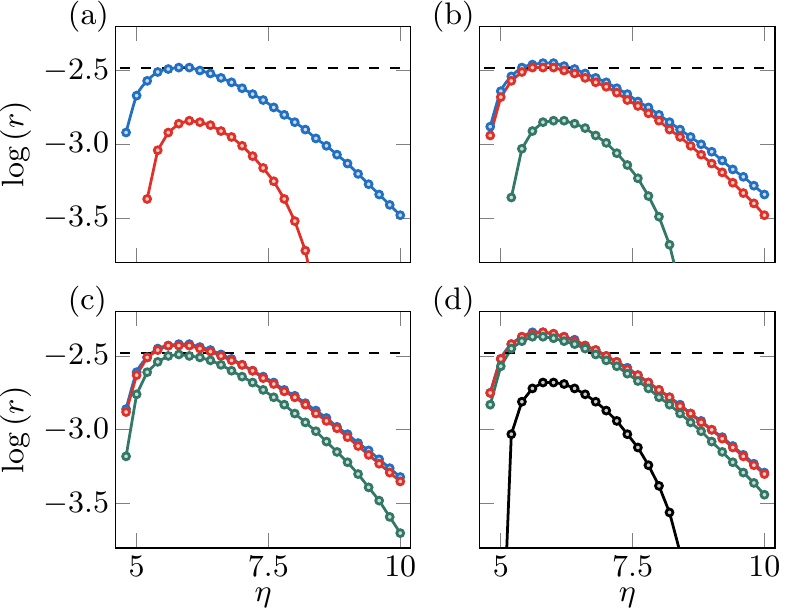}
\caption{{Maximum value of $r$ as a function of $\eta$ for $\tilde{U}_0=$ 100 (a), 10 (b), 2.5 (c) and 1.25 (d). The different lines correspond to values of $s=10^{-7.5}$ (blue), $s=10^{-6.5}$ (red), $s=10^{-5.5}$ (green), and $s=10^{-4.5}$ (black). The dashed line corresponds to the harominc trap result $r=0.0033$, for $s=0$. The lines are a guide for the eye.}\label{fig2} }
\end{center}
\end{figure} 

From the coupled equations Eq.~\eqref{eqn}-\eqref{eqv}, one can now estimate if the finite values of $r$ and $s$ hinder an efficient evaporation. Since the heating rate associated with incoherent light scattering processes gains relevance as the temperature drops, the initial evolution of the elastic collision rate is not sufficient to identify the limiting experimental parameters. Therefore, we evolve the coupled equations until the phase-space density increases by three decades in logarithmic scale. 
As an example, we show in Fig.~\ref{fig2a}, the time evolution of the normalized elastic collision rate, $\Gamma_\text{el.} = n v$, for $\tilde{U}_0 = 1.25$, $\eta = 5.8$, $s = 10^{-7}$ and different loss rate values. For $r<10^{-2.34}$ the elastic collision rate increases monotonically with time resulting in a faster evaporation. This is clear from the fact that the same gain in phase-space density is reached earlier for smaller values of $r$. For $r\approx 10^{-2.34}$ the elastic collision rate reaches a plateau as the phase-space density approaches its final value. We define this maximum value of $r$ as a threshold above which accelerated evaporation is impossible. However, this maximum does not imply a diverging elastic collision rate. 

The same analysis is performed for different values of $\tilde{U}_0$, $s$, and $\eta$ as shown in Fig.~\ref{fig2}. For the case of negligible spontaneous emission and $\tilde{U}_0 \rightarrow \infty $ we recover the well-known result $r=0.0033$ \cite{Guery-Odelin1998} \footnote{See also C. Cohen-Tannoudji lectures at Coll\`ege de France, December, 3, 1996 \href{http://www.phys.ens.fr/~cct/college-de-france/1996-97/1996-97.htm}{website}}. This result is expected as the cloud only explores the harmonic part of the dipole trap. In that limit, runaway evaporation is hindered as spontaneous emission becomes non-negligible, typically for $s\gtrsim10^{-6}$ (see Fig.~\ref{fig2}a). From an experimental point-of-view, a small value of $\tilde{U}_0$ is preferable, \textit{i.e} an initial cut-off energy approximately equal to the trap depth (see Fig.~\ref{fig2}d). In that case, the one-body loss rate condition is not too stringent for $s\lesssim10^{-4}$. Intuitively, this could be expected from Eqs.~\eqref{eqv}-\eqref{eqn}, but the dependence of $B(\eta,\, v)$, and $D(\eta,\, v)$ on the initial cut-off energy makes this assumption \textit{a priori} not obvious. With that in mind, and since $r$ and $s$ are imposed by the initial experimental conditions, the values of $\eta$ leading to an efficient evaporation can be extracted from the results shown in Fig.~\ref{fig2}.

\section{Specific example for $^{162}$Dy}

To demonstrate the relevance of the cooling protocol, we consider the specific example of $^{162}$Dy, with atoms trapped in a circularly polarized ($\sigma^{-}$) laser beam, blue detuned from the $J'=J$ optical transition at \SI{832}{\nano\meter}. For $\delta \approx 2\pi \times \SI{40}{\giga\hertz}$ the two lowest energy states $\ket{d} = \ket{J,-J}$ and $\ket{b} = \ket{J,\, -J+1 }$ have opposite polarizabilities $\alpha_{\ket{d}} = -\alpha_{\ket{b}}$ \cite{Kao2017}. The other internal states feel a strong non-linear Zeeman shift induced by the laser beam and are not coupled by radio-frequency. Finally, the ratio of real and imaginary parts of the polarizability for the two states are $\mathfrak{m}_{\ket{d}}\approx 2.8 \times 10^7 $ and $\mathfrak{m}_{\ket{b}} \approx -3.2 \times 10^6 \sim -\delta/\Gamma $, which ensures that re-pumping from state $\ket{b}$ to $\ket{d}$ is negligible over the time needed for atoms in $\ket{b}$ to leave the spatial extend of the laser beam.

We consider nominal initial experimental parameters, such as an initial temperature  $T(t=0)=\SI{30}{\micro\kelvin}$, and initial density $n_0=10^{19}\, \text{m}^{-3}$, resulting in $\Gamma_\text{el.}^{(0)} = 750 \, \text{s}^{-1}  $, $s = 5.3 \times 10^{-7}$ and initial phase space density $\mathcal{D}_0=n\lambda^3=1.6 \times 10^{-4}$. Furthermore, we assume a one-body loss rate $\Gamma_\text{loss} = 1/60\, \text{s}^{-1}$, corresponding to $r =2.2\times 10^{-5}$, and consider the case $\tilde{U}_0 = 1.25$.

\begin{figure}[t!]
\begin{center}
\includegraphics[width=\columnwidth]{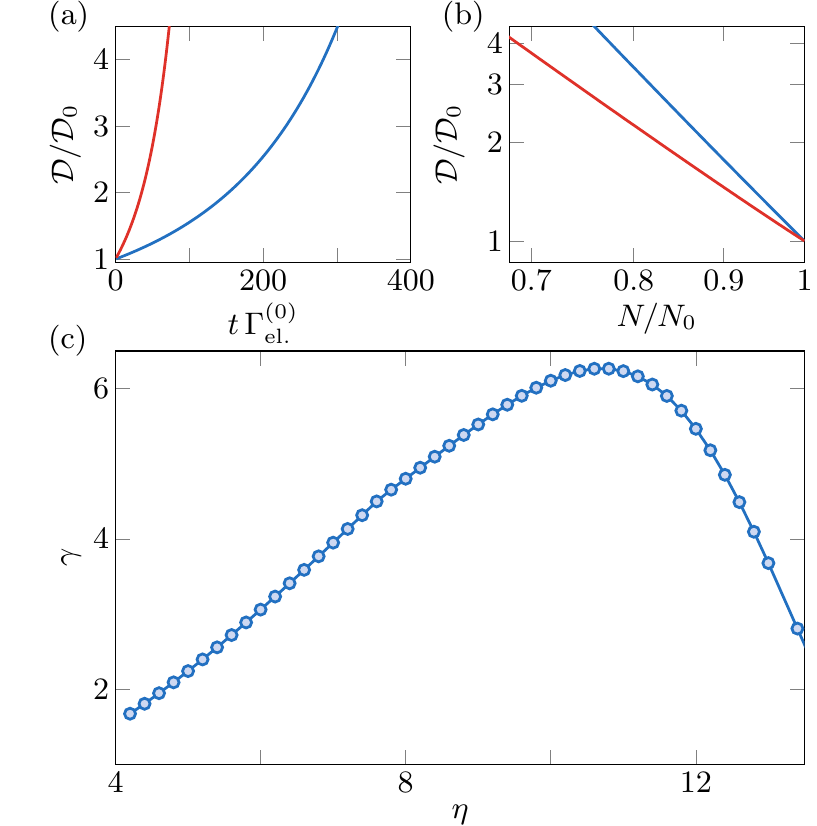}
\caption{{Results of RF evaporation in a single optical dipole trap. (a) Gain in phase-space-density as a function of time for $\eta = 7$ (red) and $\eta = 9$ (blue). (b) Logarithmic plot of phase-space-density as a function of total atom number for $\eta = 7$ (red) and $\eta =9$ (blue), resulting in $\gamma \approx 3.95$, and $5.52$, respectively. (c) Initial evaporation efficiency $\gamma$, as a function of $\eta$. The line is a guide for the eye.}\label{fig3} }
\end{center}
\end{figure} 

In Fig.~\ref{fig3}a-b, we show the gain in phase-space density for $\eta  =9$ (blue) and $\eta = 7$(red). A standard way to optimize the evaporation efficiency is ensured by maximizing the initial gain in phase space-density per particle loss, corresponding to a large value of $\gamma~=~-~{\text{dLog}(\mathcal{D} )}~/~{\text{dLog}(N)}~\big|_{t\rightarrow 0}$ 
In Fig.~\ref{fig3}c, we show the evolution of $\gamma$ as a function of $\eta$, which peaks at $\gamma \approx 6$ for $\eta\approx 10.8$. This result suggests that heating due to spontaneous emission is negligible and efficient evaporation in deep optical dipole traps is possible. In that case, the critical phase space density $\mathcal{D}_\text{c} \approx 2.6$ is reached over a time $t =4.8 \text{s} $ with a final atom number $\approx 1/4$ its initial value. One can contrast this result to another example, for instance $\eta = 6$, for which runaway evaporation also occurs. In that case, degeneracy is reached after solely $\approx \SI{0.2}{\second} $, with $4\%$ of the initial atom number. The choice of $\eta$ is therefore setup-dependent as it depends on the scientific goal and the initial experimental conditions. It is important to stress that these values are merely indicative, as the truncated Boltzmann distribution approximation is not applicable near quantum degeneracy \cite{Luiten1996a}.
\section{Discussion and conclusion}

One should note that as the elastic collision rate increases, so does the density and therefore three-body losses are enhanced. We have not considered such effect as it depends on the s-wave scattering length of the species under consideration. A related discussion of that effect can be found in Ref.~\cite{Wilkowski2010}.

In conclusion, the radio-frequency evaporation technique in an optical dipole trap, here reported, constitutes a viable path towards efficient evaporative cooling in experiments involving atomic species with a relatively narrow optical transition and where the lowest energy state has a non-zero total spin. After a general presentation of the method, we considered the specific example of Dysprosium with a laser beam near the $J'=J$ optical transition at $\lambda \approx \SI{832}{\nano\meter}$. We have shown that this technique is promising to reach quantum degeneracy in a minimalist experimental setup. Compared to standard evaporation techniques,  our method benefits from runaway evaporation which is usually unreachable for a single beam optical dipole trap.

\begin{acknowledgments}
We thank Jean Dalibard for inspiring discussions and critical reading of the manuscript. This work was supported by Grant No. ANR-20-CE30-0024.
\end{acknowledgments}

\appendix*
\section{Detailed description of Eqs.~\eqref{eqn}~-~\eqref{eqv} }
\label{appendix}

We here give the detailed expressions for the dimensionless functions $A(\eta,\, v),\, B(\eta,\, v),\, C(\eta,\, v)$, and $D(\eta,\, v)$ introduced in Eq.~\eqref{eqn}~-~\eqref{eqv}
\begin{align*}
A(\eta,\, v) & =\Gamma_\ev \pr{ \frac{(\ct_2 - \xit)(\kt- \ct)}{\ct - \xit (\eta - \ct)} -1} \\
B(\eta,\, v) & = 2 \eta \frac{\ct_2 - \xit}{\ct - \xit (\eta - \ct)} \\
C(\eta,\, v) &= \frac{\Gamma_\ev}{2} \frac{\kt - \ct}{\ct - \xit (\eta - \ct)}  \\
D(\eta,\, v) &= \eta \frac{1}{\ct - \xit (\eta - \ct)}\, , 
\end{align*}
where $\kt, \, \xit ,\, \ct ,\, \ct_2$, and $\Gamma_\ev$ are functions of $\eta$ and $v$. For $v\rightarrow 0$ and a constant, finite, $\eta$, these expressions tend to the solutions of a harmonic trap. Namely, $\Gamma_\ev = e^{-\eta} \pr{\eta - 4 R\pc{3,\eta}}$, $\kt = \eta + 1 - \frac{P\pc{5,\eta}}{P\pc{3,\eta}} \frac{e^{-\eta}}{\Gamma_\ev}$, $\xit = 3\pr{1-R\pc{3,\eta}}$, $\ct = 3R\pc{3,\eta}$. and $\ct_2 =3/2$, where $P(a,z) $ is the incomplete Gamma function, and $R(a,z) = P\pc{a+1,z}/P\pc{a,z} $ \cite{Guery-Odelin1998}.

\bibliography{../../bibfolder/ERCbib2}
\bibstyle{prsty}

\end{document}